# A GENERAL STOCHASTIC INFORMATION DIFFUSION MODEL IN SOCIAL NETWORKS BASED ON EPIDEMIC DISEASES

Hamidreza Sotoodeh[1], Farshad Safaei[2,3], Arghavan Sanei[3] and Elahe Daei[1]

[1] Department of Computer Engineering, Qazvin Islamic Azad University, Qazvin, IRAN
{hr.sotoodeh, e.daei}@qiau.ac.ir
[2] School of Computer Science, Institute for Research in Fundamental Sciences (IPM), P.o.Box 19395-5746, Tehran, IRAN
safaei@ipm.ir
[3] Faculty of ECE, Shahid Beheshti University G.C., Evin 1983963113, Tehran, IRAN
f_safaei@sbu.ac.ir, a.sanei@mail.sbu.ac.ir

## ABSTRACT

*Social networks are an important infrastructure for information, viruses and innovations propagation. Since users' behavior has influenced by other users' activity, some groups of people would be made regard to similarity of users' interests. On the other hand, dealing with many events in real worlds, can be justified in social networks; spreading disease is one instance of them. People's manner and infection severity are more important parameters in dissemination of diseases. Both of these reasons derive, whether the diffusion leads to an epidemic or not. SIRS is a hybrid model of SIR and SIS disease models to spread contamination. A person in this model can be returned to susceptible state after it removed. According to communities which are established on the social network, we use the compartmental type of SIRS model. During this paper, a general compartmental information diffusion model would be proposed and extracted some of the beneficial parameters to analyze our model. To adapt our model to realistic behaviors, we use Markovian model, which would be helpful to create a stochastic manner of the proposed model. In the case of random model, we can calculate probabilities of transaction between states and predicting value of each state. The comparison between two mode of the model shows that, the prediction of population would be verified in each state.*

## 1. INTRODUCTION

In recent decades, networks provide an infrastructure that economic, social and other essential revenues are depending on. They can form the physical backbones such as: transportation networks (convey vehicle flows from sources to destinations), construction and logistic ones (provide transforming the row material and presenting the ultimate products), electricity and power grid ones (consign required fuels) and Internet ones [1] (provides global public accesses and communications). These structures lead to thousands of jobs, social, politics, economics, and other activities. Moreover, complex physical networks are also established such networks, like financial, social and knowledge networks, and those ones are under development as smart grid [2].

*Social network* has declared as a structure, that its entities can communicate with each other through various ways. These entities denote as users [3]. So, users play the main role in construction of social networks. Since, the social networks are abstraction of a real world; many social phenomena can be modeled at the level of social networks [4]. Dealing with all kind of diseases among population of a society, could be one major issue of them. Communication between infectious and susceptible users makes dissemination of these diseases. Outbreak of a disease and also, how people behave in the face of contagious can be the reason of modeling information diffusion in social networks [5].

*Information diffusion* is a general concept and is defined as all process of propagation, which doesn't rely on the nature of things to publish. Recently, the diffusion of innovations and diseases over social networks has been considered [6]. These models assume users in a social network are influenced by the others, in other words, they model processes of information cascades [7]. This process makes an overlay network on the social or information network. Context network and power of data influence effect on, how data spreads over the network [7, 8].

Many mathematical studies have been done on *disease diffusion*, assuming population in a society has totally homogeneous structure (i.e., individuals behave exactly the same as each other) [9, 10]. This assumption allows writing easier the nonlinear differential equations, which describe the individual's behavior. But, this assumption is not realistic; because, the structures and features of individuals are not the same as each one [11]. For example, they don't have the same capability of transferring and caching diseases. Therefore, population can be divided into some groups, that individuals in each community have similar abilities and structures; however, they have different capabilities in comparison with the other communities. This model refers to the *compartmental epidemic models*, which treat nearly real treatments [12]. By the thanks of difference equations, we can obtain the number of infected individual as a function of time. Also, it can be obtained the size of diffusion and be discussed about, whether the epidemic has occurred or not?

Furthermore, epidemic behavior usually declares via a transition phase, which takes place whenever it can be jumped from epidemicless state to a condition which contains that. Basic reproduction number ( $R_0$ ) is a trivial concept in epidemic disease and determines this mutation from these two states [10]. This parameter has been defined as a *threshold*, which if $R_0 > 1$, then the spreading of the contamination can be occurred through the infected individual. On the other hand, if $R_0 < 1$, then contamination cannot outbreak among all population, disease will be disappeared. So, the epidemic doesn't occur [13].

Generally, epidemic models divided into two deterministic and stochastic categories [14]. The most important concern of deterministic models is their simplicity, which proposed in communities for large population. Usually in these models, we can find some questions as [15]:

- Will the entrance of a disease lead to an epidemic manner? Or when does a disease go into the epidemic?
- How many people are affected? Or what is the influence of immunization (vaccination) to the part of a community that has been incurred?
- ...

However, as mentioned earlier, the oblivious fact is people are in different level of infections. If a group of people connected to one infected person, all those population could be affected. By grouping of individuals with their capacity, there are still some questions [15]:

- What is the probability, that individuals get the infection per each group?
- Or sometimes, the situations may arise that, how we can predict the probability of disease dissemination in a large population?

Stochastic methods are usually suitable to response these questions. However, these ones have more complication than deterministic methods. Better solution is presenting a stochastic model based on deterministic one. Three approaches have been proposed of these methods [16], two of them are related to DTMC[1] and CTMC[2] Markovian models and the third one suggests using SDE[3] methods.

---

[1] Discrete Time Markovian Chain.
[2] Continuous Time Markovian Chain.

In this paper, our goal is present a general model of information diffusion, which is based on epidemic diseases. In fact, this model actually is a result of developing SIRS deterministic model and including compartmental assumption. We aim to adapt our proposed deterministic model to stochastic one by using discrete-time Markovian model. It should be expressed that, the graph structure impacts to process of diffusion definitely; however, in this paper this affection is not be considered and assumed all structures have the same affection. The outline of the paper is as follows: we first introduce information diffusion models and related work of them especially about epidemic models in section 2; then, propose a general model to diffusion based on SIRS deterministic epidemic model in section 3; moreover, we get into the analytical model description and obtaining important threshold measurement for our model during this section. Another major aim of this section is leading our deterministic model to a stochastic one by the help of DTMC model. Finally, in section 4, we bring experimental results, in this section we are going to validate our measures that have gotten from our model.

## 2. RELATED WORK

 Researchers usually consider two approaches to model information diffusion, dependent and independent of graphs. Graph-based approaches focus on topology and graph structure to investigate of their impact on spreading processes. Two of the most important diffusion models in this class are *Linear Threshold* (LT) [8] and *Independent Cascade*(IC) [17] models. These are based on a directed graph, where each node can be activated or inactivated (i.e., informed or uninformed). The IC model needs probability to be assigned to each edge, whereas the LT needs an influence degree to be declared on each edge and an influence threshold for each node. Both of the models do the diffusion process iteratively in a synchronous way along a discrete-time, starting from a set of initially activated nodes. In the IC, newly activated nodes try to activate their neighbors with the probability defined on their edges. This activity has been done for per iteration. On the other hand, in LT, the active nodes join to activate sets by their activated neighbors when the sum of the influence degrees goes over their own influence threshold. This event is done at per iteration of this process. Successful activations are effective at the next iteration. In both models, the diffusion end while there is no neighboring node can be contacted.  These two models rely on two different perspectives: IC refers to sender-centric and LT is receiver-centric approach. With the sake of both these models have the inconvenience to proceed in a synchronous way along a discrete time, which doesn't suitable in real social networks. In order to establish more consistency on real networks, can referred to *ASIC* and *ASLT* models [18], which are asynchronous extension of these models. These mechanisms use continues time approach and for this, each edge would be equipped a time-delay parameter.

In the case of independent graph models, there isn't any assumption about graph structure and topology impaction on the diffusion. These models have been mainly developed to model epidemic processes. Nodes are organized to various classes (i.e. states) and focused on evaluation of proportion of nodes in each state. *SIR* and *SIS* are two famous instant of these models [9, 19]. Acronyms "S" is for susceptible state, "I" declare infectious (informed person) state. In both models, "S" can exchange their state to "I" with a constant rate (e.g. $\alpha$). Then, in SIS model nodes can make a transition to "S" state again with the constant rate (e.g. $\beta$). In case of SIR model nodes, can switch to "R" (stand for Removed/Recovery) situation permanently. The percent proportion of population determine by the help of difference equations. Both models assume that every node has the same probability to be connected to another; thus connections inside the population are made at random. But, the topology of the nodes relations is very important in social network; as a result, the assumptions made by these models are unrealistic.

A good survey regarding to analysis, developing, vaccination and difference equations of populations growth has been done in [20], and this research is based on *SIR* and *SIS*, two famous epidemic models.

---

[3] Stochastic Differential Equation.

But, as mentioned earlier, in these models assume, whole the network as a homogeneous and all the individuals is in equal statuses (i.e. similar node degree distributions, equal infection probability and …). In case of these models, each person has the same relation to another one and contagion rate is determined by density of infected peoples, which mean dependent to the number of infected individuals.

The virus propagation on homogeneous networks with epidemic models has been analyzed by Kephart and Wiht [13]. One of the important things that could be obvious is real networks including social networks; router and AS[4] networks follow a power low structure instead. One study has done on non-homogeneous models and in there population divided into separate communities with each own features [13]. These models are part of compartmental models. The result can be deduced from [13], is using one important theorem, which declares the relation between threshold parameter and individual relationships as matrix Eigenvalues.

Following the work of [20], many epidemic models have developed till now, which each one can be useful as their applications. A good reference that surveyed these models could be found in [21]. The compartmental models have been developed as a good manner in [22]; the authors' work is based on SIR model and susceptible individuals decomposed to different groups with the same properties. In this model immunized people will permanently vaccinate and would not return to population. This proposed model, have been developed afterwards and been created a new differential model, which immunized people after a time can join to susceptible group [23]. Moreover, there are two time periods, first one, latent time period to appearance symptoms of disease that a person infected, and the second one is to start of transmission. Then, they present the extended SEIRS model for virus propagation on computer networks. Given the deterministic models represent the equations of population changes as well, but random behavior of users in real social networks, cause deterministic models to stochastic ones.

One of the rich study have focused on three methods to obtain stochastic models, according to deterministic models [16]. Two of them related to *DTMC* and *CTMC* Markovian models and the third one is suggested to use SDE methods for this aim. Epidemic threshold is one of the important parameters, which get considered in all studies. A survey has been investigated on this parameter as their major applications [24]. Pastor-Satrras and Vespignani [25] have studied virus propagation on stochastic networks with power low distribution structure. In such networks, have been showed that, threshold has a trivial value meaning; that even an agent with extremely low infectivity could be propagated and stayed on them. "Mean-field" approach is used by them, where all graphs should have similar behavior in terms of viral propagation in recent work; Castellano and Pastor-Satorras [26] empirically argue that, some special family of random power-low graphs have a non-vanishing threshold in SIR model over the limitation of infinite size, but provide no theoretical justification. Newman [10, 27] studied threshold for multiple competing viruses on special random graphs, accordingly mapping SIR model to a percolation problem on a network. The threshold for SIS model on arbitrary undirected networks has been given by Chakrabarti et al. [28] and Ganesh et al.; finally, Parkash et al. [29], focuses on the arbitrary virus propagation models on arbitrary, real graphs.

## 3. THE PROPOSED DIFFUSION MODEL

The model, which we will propose for information diffusion in a social network, inspired of *SIRS* deterministic model of epidemic disease [30]. The person who is *susceptible* (i.e. waiting to get information) denoted by "S", and the person who gets a new information will run into *active*, "A" state. Finally, rest of the population belongs to *deactivate* state would be appeared with "D". Deactivated users may return to networks after period of a time. A person who has returned, count for a susceptible one, since it can be influenced another reason and transfers to the same or different infection states. According to the nature of users' behavior in social networks, the proposed model inclined toward a compartmental model. This means that, each state can be divided into several groups regarding to users' features. All of the people in a group are similar to each other, whereas people in different groups are distinct. Figure 1 shows the state diagram of this model in general. Before analyzing the model in details, it is necessary to explain some assumptions, that the model is based on them.

---

[4] Autonomous Systems

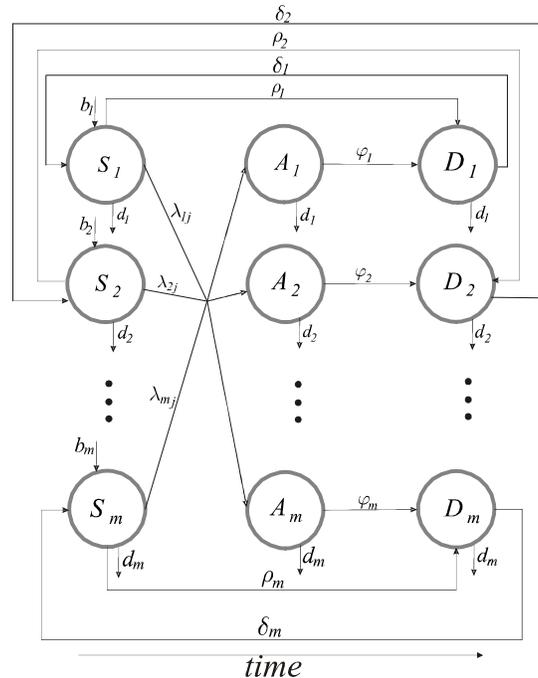

**Figure 1.** A visual representation of our model for $m$ compartments in each state. Here is a probability of transmission a susceptible user to any of active state (who is informed) or deactivate one. Also, there would be possible that move to a deactivate state after it informed. All the arcs are included with the probability (rates), which these transmissions would occur.

## 3.1. MODEL ASSUMPTIONS

The main assumptions of this model can be summarized as follows:

1. Total population size is constant, $N$ ( $N = \sum_{i=1}^{m} S_i(t) + \sum_{i=1}^{m} A_i(t) + \sum_{i=1}^{m} D_i(t)$ ). This assumes that the space is closed world, which have been considered internal constraints and relationships only. It would be impossible that information come from the external source in network.

2. Each one of the *susceptible*, *active* and *deactivate* states divided into $m$ different groups according to the user's features.

3. Any new node (i.e. user) has been added into the network; it would be susceptible and is belong to one of the $s_i$ states. The rate of joining to each of $s_i$ groups would be constant $b_i$ .

4. The natural death rate of the nodes for each group would be constant $d_i$ . This rate explains the permanent disjoint according to the network; however, at deactivate state $D_i$, which exists a probability to return to the network.

5. Different users of each groups, $s_i$ can migrate to one of the active groups, $A_j$ with the rate $\lambda_{ij}$ .

$\lambda_{ij}$ stands for transfer rate from $S_i$ to $A_j$ state ( $i \in S \ni 1 \le i \le m, j \in A \ni 1 \le j \le m$ ).

6. Infectious period for each user of each $A_i$ group is constant.

7. Transferring, between susceptible $s_i$ and deactivate groups is done with rate $\rho_i$. Note that, deactivate state is equal for all population, but people can return to the same state that they had activated before. So, dividing this state to $m$ groups, allow the transmission to analogous susceptible groups.

8. Each users of $A_i$ can inactivate themselves for Different reasons. This change can be done with rate, $\varphi_i$ for any $A_i$ state to $D_i$ .

9. Users which have gone to inactive state after return to the network would be in susceptible state again. While users were being deactivated, they are away from new information and don't get any information. So, as they return from the deactivate state, cannot join to active groups and should return to the status, as they had been inactivated before ( $S_i$ ). This migration is done with rate $\delta_i$ .

Most mathematical models, which have developed in this area, are based on above assumptions [9, 10, 12, 15, 22, and 23].

## 3.2 Problem of Interest

Before we get into the detail of our approach, it is useful to notify some criteria and metrics, that we are eager to derive from the proposed model and is used in whole of the paper. One of the important measures per epidemic diffusion processes is a threshold value, which is a criterion for determining, whether the epidemic take place or not. Based on the diffusion type, this parameter can give distinct value with different parameters. Firstly, we can calculate this measure for our model and determine when an epidemic can occur following this model. Then, specify how this value effects on diffusion process. To better adaption the model to reality, we use DTMC Markovian model, since this model explains better random users' behavior. Calculating probability transition and then expected number of users in each state of epidemic can be attractive. Also, once information (beneficial or no beneficial) extinct in network, determining the end of diffusion process, "when does it happen?" would be another motivation of this paper. In the following, we investigate to calculate these parameters; but, before we will consider an analytical description of the proposed model.

## 3.3. Analytical Description of the Proposed Model

In this section, we'll investigate mathematical description for our proposed model. User's treatment can be modeled by difference equations, which made on getting information. According to assumptions in section 3.1, following equations could be deduced.

$$
\begin{cases}
S_i^{'}(t) = b_i - \sum_{j=1}^{m} \lambda_{ij} S_i(t) - (d_i + \rho_i) S_i(t) + \delta_i D_i(t) \\[2mm]
A_i^{'}(t) = \sum_{j=1}^{m} \lambda_{ij} S_i(t) - (d_i + \varphi_i) A_i(t) \\[2mm]
D_i^{'}(t) = \varphi_i A_i(t) + \rho_i S_i(t) - (d_i + \delta_i) D_i(t)
\end{cases}
\tag{1}
$$

Based on above equations, $S_i^{'}(t)$, $A_i^{'}(t)$ and $D_i^{'}(t)$ denote variations of susceptible of information, informed users and removed ones, respectively. Also, the number of susceptible, informed person and removed users is according to time. Once one user gets information, it can move from its own state. Entrance and exit of its state is determined with sign (+) and (–), respectively.

Regarding to Assumption 5 of the model $\lambda_{ij}$ is stand for the rate between $S_i$ and $A_j$ states. However, this parameter relates to two other measures: susceptibility of each user in per groups of susceptible $\varepsilon_i$ and infectiousness of each user in any infection groups $\gamma_j$.

Also, as mentioned earlier, another important measure, that impact to this parameter is network structures. This measure denotes relationship between users in a network. Certainly, different graphs have distinct affect. We have to impact this measure as α. In other words, capability of getting data (infection) relates to rate of the relationships and number of the infected users, who is involved in relationships.

As a result, $\lambda_{ij}$ should be defined as how that three mentioned measures, $\varepsilon_i$, $\gamma_j$ and $\alpha$ can be effective to calculate it. Values of these parameters, determine as random, which was assigned to each their groups. For calculating $\alpha$, relations between users in a network should be considered. However, in this paper we ignore this influence and just have used a random value for this parameter. As a mathematically form, we can use Equation (2) to calculate $\lambda_{ij}$ :

$$
\lambda_{ij} = \alpha \varepsilon_i \gamma_j \frac{A_j}{N},
\tag{2}
$$

where $\frac{A_j}{N}$, denotes the percent of informed users of group $j$ to the total users in network. In other words, it explains the fraction of users with transmission strength $\gamma_j$, can inform susceptible users of group $i$ with susceptibility $\varepsilon_i$. According to user's behavior, one of them can disjoin from the network and would return again. So, the parameter $\alpha$ should be updated based on $\varphi_i$, $\delta_i$ and $b_i$ parameters. This action can be done with considering the least values of them $min(\varphi_i, \delta_i, b_i)$.

Regarding to movement of users, changes can accrue in the system; so whenever system is at stable state, there would be no more changes in the states ($S_i^{'}(t), A_i^{'}(t), D_i^{'}(t) \sim 0$). Since, the population of these states would be based on information, whether gets epidemic or not. So, the threshold parameter $R_0$ plays an important role on population size of each group. These sizes can be deduced in stable state from Equation (3).

$$\begin{cases} (1)\, R_0 < 1 \Rightarrow \\ \quad \lim_{t \to \infty} (S_i(t), A_i(t), D_i(t)) = (S_i^*(\bar{t}), 0, D_i^*(\bar{t})) = \\ \qquad (\dfrac{b_i d_i + \delta_i}{d_i(d_i + \delta_i + \rho_i)}, 0, \dfrac{b_i(\rho_i + d_i)}{d_i(d_i + \delta_i + \rho_i)}), \\ (2)\, R_0 > 1,\, A_i > 0 \Rightarrow \\ \quad \lim_{t \to \infty} (S_i(t), A_i(t), D_i(t)) = (Si^*(\bar{t}), Ai^*(\bar{t}), Di^*(\bar{t})) = \\ \qquad (\dfrac{\left[\dfrac{d_i + \delta_i}{d_i}\right] b_i - A_i(\varphi_i + d_i + \delta_i)}{\rho_i + d_i + \delta_i}, \dfrac{\left[\dfrac{d_i \delta_i}{d_i}\right] b_i}{(\varphi_i + d_i + \delta_i) + \dfrac{(\varphi_i + d_i)(\rho_i + d_i + \delta_i)}{\sum_{j=1}^{m} \lambda_{ij} A_j}}, \dfrac{\varphi_i A_i + \rho_i S_i}{d_i + \delta_i}). \end{cases} \tag{3}$$

- Case 1: occurs when new information on network doesn't get epidemic in stable state, or we are in disease-free equilibrium ( $A_i \to 0$ ).
- Case 2: whenever happen that propagated information on network is pervasive, or we are in endemic equilibrium ( $A_i > 0$ ).

In case 2, the value of $A_i^*$ is calculated in parametric form. But, for $S_i^*$ due to lake of space, we ignore to write the parametric value, as well as for calculating $D_i^*$. In the following, we calculate the threshold parameter to our proposed model. However, before that, we are going to define this parameter in detail and introduce how to get it in compartmental models briefly.

### 3.3.1. Threshold Parameter to determine the Extent of the Epidemic Dissemination of Information

One of the most important concerns about any infectious disease is its ability to invade a population. Many epidemiological models have a disease free equilibrium (DFE) at which, the population remains in the absence of disease. These models usually have a threshold parameter, known as the basic reproduction number $R_0$; such that, if $R_0 < 1$, then the DFE is locally asymptotically stable, and the disease cannot invade the population; because, on average an infected individual produces less than one, new infected individual over the course of its infectious period. Conversely, if $R_0 > 1$ then each infected individual produces, on average more than one new infection, and the disease can invade the population. So, the DFE is unstable [22]. A definition for this threshold can be defined, as below.

**Definition 1 [31]**: The *basic reproduction number denoted by $R_0$ is 'the expected number of secondary cases produces, in a completely susceptible population, by a typical infective individual.*

In the case of a single infected compartment, $R_0$ is simply the product of the infection rate and the mean duration of the infection. However, for more complicated models with several infected compartments this simple heuristic definition of $R_0$ is insufficient. Further, the general basic reproduction number can be defined as the number of new infectious produced by a typical infective individual in a population at a DFE.

There exist different approaches to calculate this parameter. Table 1 shows these methods with their references. According to the proposed model belongs to compartmental models; the next generation matrix is a suitable method to obtain threshold value.

**Table 1.** Different Methods to calculate threshold parameter, $R_0$ in the literatures.

| Methods | Reference |
|---|---|
| Next-generation Operator | [32-34] |
| Local stability of disease-free equilibrium | [19, 39-45] |
| Existence of an endemic equilibrium | [40,41] |
| Multiple criteria | [42-44] |

### 3.3.1.1. A general Compartmental Model

Let consider a heterogeneous population, that individuals are not distinguishable from each other; according to age, behavior, spectral position and … So, this population can be grouped to *n* subpopulation. In this section, it would be explained one general epidemic model for this population, which presented by van den Driessche and Watmough [31]. Although these models are based on a continuous-time model, we focus on the discrete –time of that and later will adapt it to our model.

Consider a population vector $x = (x_1, ..., x_n) \mid x_i \geq 0$, that defines the number of individuals in each of *n* compartment and let the first *m* of these compartments correspond to infected conditions. To calculate the value of $R_0$, determining a new infection is more important than other changes in the same population.

Assume $\mathcal{F}_i(x)$ is the rate of appearance of new infections in compartment $i$, $\mathcal{V}_i^+(x)$ represent the rate of movement of individuals into compartment $i$ by means of other infection; $\mathcal{V}_i^-(x)$ is the rate of removal of individuals from compartment $i$ by any means. Note that, distinction between terms included in $\mathcal{F}_i(x)$ and $\mathcal{V}_i^+(x)$ is not mathematical, but biological; this distinction impacts the computation of $R_0$.

We can formulate a difference equation model of this process as follows:

$$x_i^{'} = f_i(x) = \mathcal{F}_i(x) + \underbrace{\mathcal{V}_i^+(x) - \mathcal{V}_i^-(x)}_{-\mathcal{V}_i(x)}, \tag{4}$$

The only restrictions placed on the form of the functions, are given by the following assumptions. These hypotheses don't effect on the behavior of functions though these are biological assumptions, which impact on the model treatments.

**1**. Since that, each function determines the direction of individual transmission, all of them is nonnegative. Or *if $x > 0$, then $\mathcal{F}_i(x), \mathcal{V}_i^+(x), \mathcal{V}_i^-(x) \geq 0$, $i \in \{1, ..., n\}$.*

**2**. If one of the subpopulation is empty, there will be impossible to transmit any of them. Or *if $x_i = 0$ then $\mathcal{F}_i(x) = 0$ and $\mathcal{V}_i^+(x) = 0, i \in \{1, .., m\}$.*

**3**. Outbreak and spread of disease from the non-infected subpopulation could be impossible and equal to zero. Or
*$\mathcal{F}_i(x) = 0$ for $i > m$.*

**4**. People cannot get out of the subpopulation over than its capacity .Or
*$\mathcal{V}_i^-(x) \leq x_i$.*

As mentioned in previous sections, if the basic reproduction number $R_0$ is less than one, then the DFE is locally asymptotically stable, and also the disease cannot spread easily. However, whenever a population vector(x) can be called as DFE, the first m compartments of x are zero (i.e., $\{x \geq 0 \mid x_i = 0, i = 1,...,m\}$ ).

While all of the Eigenvalues of the Jacobian matrix of the function at the equilibrium $x$ ( $J = -D\mathcal{V}(x)$ ), have modulus less than one, we can say the disease free population dynamics is locally stable. This condition is based on the derivation of $f$ near (or at) a DFE. For our purpose, we define a DFE of (4) to be a stable equilibrium solution of disease free model. Note that, it is not necessary assume the model has a unique DFE. Consider a population near the DFE $x_0$. If the population remains near the DFE (i.e. if the introduction of a few infective individuals does not result in an epidemic) then the population will return into the DFE, according to the linearized system

$$x^{'} = Df(x_0)(x - x_0),\tag{5}$$

where, $Df(x_0)$ is the Jacobian matrix (i.e. $\left(\dfrac{\delta f_i}{\delta x_j}\right)$ that evaluated at the DFE, $x_0$ ). In stability analysis, can be used the Eigenvalues of the Jacobian matrix evaluated at an equilibrium point, to determine the nature of that equilibrium. If all of the Eigenvalues are negative, the equilibrium is stable. Because, we restrict our attention to systems in the absence of new infection, so, the stability of DFE determined by the Eigenvalues of $J = -D\mathcal{V}(x)$.

### 3.3.1.2 CALCULATING THE THRESHOLD PARAMETER IN THE PROPOSED MODEL

Before going into details of calculating this parameter in our model, first we explain a theorem to obtain this parameter for compartmental model, which come its description earlier. The proof of this theorem can be found in [31, 45].

**Theorem 1 [45]:** *Let x be a DFE and define the* $m \times m$ *matrices* $F = \{f_{ij}\}$ *and* $V = \{v_{ij}\}$ *as:*

$$f_{ij} = \frac{d\mathcal{F}_i}{dx_j}|_x, v_{ij} = \frac{d\mathcal{V}_i}{dx_j}|_x \ i, j \in \{1,...,m\},\tag{6}$$

*the next-generation matrix is given by* $K = FV^{-1}$, *so* $R_0 = \rho\left(FV^{-1}\right)$, *where* $\rho$ *denote the spectral radius of matrix K. The DFE x is locally asymptotically stable if and only if the spectral radius of the Jacobian,* $\rho(I + F - V)$, *is less than 1, which in turn occurs if and only if is* $R_0$ *less than 1.*

To interpret the entries of $FV^{-1}$ and develop a meaningful definition of $R_0$, consider the fate of an infected individual into compartment $k$ of a disease free population. The $(j, k)$ entry of $V^{-1}$ is the average period of time this individual spends in compartment $j$ during its lifetime. Assuming that the population remains near the DFE and barring reinfection. The $(i, j)$ entry of $F$ is the rate of which infected individuals in compartment $j$ produce new infections in compartment $i$. Hence, the $(i, k)$ entry of the product $FV^{-1}$ is the expected number of new infectious in compartment $i$ produced by the infected individual originally introduces into compartment $k$. Following Diekmann et al. [32], we call the next generation matrix for the model and set $R_0 = \rho\left(FV^{-1}\right)$.

Now, we would like to calculate $R_0$ by decomposing the system of ordinary differential equation into the new information and transfer of user as below.

$$
\left\{
\begin{aligned}
\mathcal{F} &= \begin{pmatrix} 0 \\ \sum_{j=1}^{m} \lambda_{ij}(x) S_i(t) \\ 0 \end{pmatrix}. \\[2em]
\mathcal{V} &= \begin{pmatrix} -b_i + \sum_{j=1}^{m} \lambda_{ij}(x) S_i(t) + (d_i + \rho_i) S_i(t) - \delta_i D_i(t) \\ (d_i + \varphi_i) A_i(t) \\ -\varphi_i A_i(t) + \rho_i S_i(t) + (d_i + \delta_i) D_i(t) \end{pmatrix}.
\end{aligned}
\right.
\tag{7}
$$

Further, with calculating the Jacobian at DFE we have:

$$
\left\{
\begin{aligned}
F &= \begin{pmatrix} \lambda_{11}(x_0) S_1^* & \cdots & \lambda_{1m}(x_0) S_1^* \\ \vdots & \ddots & \vdots \\ \lambda_{m1}(x_0) S_m^* & \cdots & \lambda_{mm}(x_0) S_m^* \end{pmatrix} = \left[ \lambda_{ij}(x_0) S_i^* \right]_{ij}, \\
&and \\
V &= \begin{pmatrix} (d_1 + \varphi_1)\phi_{11} & \cdots & (d_1 + \varphi_1)\phi_{1m} \\ \vdots & \ddots & \vdots \\ (d_m + \varphi_m)\phi_{m1} & \cdots & (d_m + \varphi_m)\phi_{mm} \end{pmatrix} = \left[ (d_i + \varphi_i)\phi_{ij} \right]_{ij}.
\end{aligned}
\right.
\tag{8}
$$

Where in the above equations $S_i^*$ come from the system at the DFE, which we calculated in Equation (3). We can rewrite stable condition as $\left( \dfrac{b_i d_i + \delta_i}{d_i (d_i + \delta_i + \rho_i)}, 0, \dfrac{b_i (\rho_i + d_i)}{d_i (d_i + \delta_i + \rho_i)} \right)$. Also, about $\phi_{ij}$, we have:

$$
\phi_{ij} = \begin{cases} 0 & if \ \ i \neq j \\ 1 & if \ \ i = j \end{cases}.
\tag{9}
$$

Thus, we can calculate next generation matrix regard to its spectral radius as a reproduction number:

$$
FV^{-1} = \left[ \frac{\lambda_{ij}(x_0) S_i^*}{(d_i + \varphi_i)} \right]_{ij} \Rightarrow R_0 = \rho\left( FV^{-1} \right).
\tag{10}
$$

if we choose $\lambda_{ij} = \omega_i \times \eta_j$ [46], which the next generation matrix has rank one, we can write $R_0$ as:

$$
R_0 = \sum_{i=1}^{m} \frac{\omega_i(x_0) \eta_i(x_0) S_i^*}{(d_i + \varphi_i)},
\tag{11}
$$

where $\omega_i$ and $\eta_j$ are constant. In this situation, spectral radiance can be calculated by summation diagonal of $\left(FV^{-1}\right)$ matrix.

## 3.4. The Transaction Probability Calculation

The model that we present for information diffusion in section 3.2 is based on deterministic model. It should be noticed that, modeling and analyzing deterministic modes would be easier than stochastic ones. However, these cases are away from the reality. So, we are going to use *DTMC* [47], to justify our model to a stochastic manner, since it adapts to real user's behavior more than deterministic case.

To calculate probability transactions with the help of Markovian model, we should put on some hypothesizes before using Markovian model.

1. Markov property [47]: this means, to predict the future state we just consider present state and ignore the previous states. We can show this property at mathematical format as:

   if $\left\{X(t), t \geq 0\right\}$ denotes a set of stochastic variables, where $X(t)$ is based-on a Markov process, then for all values of $t_1 < t_2 < ... < t_m < t_{m+1}$, we can define following equation:

   $$P\left\{X(t_{m+1}) \leq x \mid X(t_1) = x_1, X(t_2) = x_2, ...., X(t_m) = x_m\right\}$$
   $$= P\left\{X(t_{m+1}) \leq x \mid X(t_m) = x_m\right\}. \tag{12}$$

2. We define three stochastic variables to determine the number of users in susceptible, active and deactivate states at time t.

   $$\begin{cases} \mathcal{S}_i(t), \mathcal{A}_j(t), \mathcal{D}_k(t) \in \left\{0, 1, 2, ...., N\right\}, i, j, k = 1..m \\ \sum_{k=1}^{m} \mathcal{D}_k(t) = N - \sum_{i=1}^{m} \mathcal{S}_i(t) + \sum_{j=1}^{m} \mathcal{A}_j(t), i, j, k = 1..m \end{cases} \tag{13}$$

3. Time in DTMC model should be divided into small time units, that one event occur only per epoch $(t \in \left\{0, \Delta t, 2\Delta t, ...\right\})$ [16].

According to Assumption 2, we can consider two stochastic variables, $\mathcal{S}_i(t), \mathcal{A}_i(t)$, the random variable $\mathcal{D}_i(t)$ can be calculated as $\sum_{k=1}^{m} \mathcal{D}_k(t) = N - \sum_{i=1}^{m} \mathcal{S}_i(t) + \sum_{j=1}^{m} \mathcal{A}_j(t), i, j, k = 1..m$. So, we don't have to consider it in the computations. As a result, bivariate process $\left\{\left(\mathcal{S}_i(t), \mathcal{A}_j(t)\right)\right\}_{t=0}^{\infty}$, has a joint probability function given by:

$$p_{(s_i, a_j)}(t) = Prob\left\{\left(\mathcal{S}_i(t) = s_i, \mathcal{A}_j(t) = a_j\right)\right\} \tag{14}$$

This bivariate process has the Markov property and is time-homogeneous. Transition probabilities can be defined based on the assumptions of the deterministic formulation. First, assume that, Δt can be chosen small sufficiently, that at most one change in state occurs during the time interval $\Delta t$. The transition probabilities are denoted as follows:

$$p_{(s_i+k,a_j+l),(s_i,a_j)}(\Delta t) = Prob\left\{\left(\Delta\mathscr{S}_i,\Delta\mathbb{A}_j\right) = (k,l) \mid \left(\mathscr{S}_i(t),\mathbb{A}_j(t)\right) = (s_i,a_j)\right\}.$$

$$i,j = 1..m, \quad \Delta\mathscr{S}_i = \mathscr{S}_i(t+\Delta t) - \mathscr{S}_i(t) \tag{15}$$

Hence,

$$p_{(s_i+k,a_j+l),(s_i,a_j)}(\Delta t)_{b_i=0} =$$

$$\begin{cases}
\sum_{j=1}^{m}\lambda_{ij}s_i\Delta t = \theta_1 & (k,l) = (-1,1) \\
a_j(\varphi_i + d_j)\Delta t = \theta_2 & (k,l) = (0,-1) \\
\delta_i(N - s_i - a_j)\Delta t = \theta_3 & (k,l) = (1,0) \\
(d_i + \rho_i)s_i\Delta t = \theta_4 & (k,l) = (-1,0) \\
1 - (\theta_1 + \theta_2 + \theta_3 + \theta_4) = \theta_5 & (k,l) = (0,0) \\
0 & Otherwise
\end{cases} \tag{16}$$

According to Equation (16), probability of transitions calculated per time step, $\Delta t$. in this time, just one movement can be taken between states because $\Delta t$ is sufficiently, small which only one event can take place. For example, the probability of $(k,l) = (-1,1)$ means that, for each state $i$, only one person can migrate from state $s_i$ to $a_j$. This probability has calculated as a specific state $i$, so we can transfer to one of $j$ state from $a$ states (i.e. $\sum_{j=1}^{m}\lambda_{ij}s_i\Delta t$, which equals to $\theta_1$ symbol). All probabilities have been calculated similar to this one. Moreover, in Equation (16) assumed that, birth rates at each susceptible states, are zero and sum of the all probabilities lead to one, so this avoids additional computations.

The time step $\Delta t$ must be chosen sufficiently small, such that each of the transition probabilities lie in the interval [0, 1]. The sake of these states is now ordered pairs, the transition matrix become more complex and its form depends on, how the states $(s_i,a_j)$ are ordered; moreover, this matrix is a Trier dimension. The column of this matrix declares as $(s_i,a_j)$ and the rows define $(s_i+k,a_j+l)$ pairs. Third dimension would be applied for $(i,j)$ pair, which $1 \le i, j \le m$. so, we have avoided showing this matrix here. However, applying the Markov property, the difference equation satisfied by the probability $P_{(s_i,a_j)}(t+\Delta t)$ and can be expressed in terms of the transition probabilities.

$$p_{(s_i,a_j)}(t+\Delta t) = p_{(s_i+1,a_j-1)}(t)\sum_{j=1}^{m}\lambda_{ij}^{a_j-1}(s_i+1)\Delta t + p_{(s_i,a_j+1)}(t)(a_j+1)(\varphi_i+d_j)\Delta t +$$

$$p_{(s_i-1,a_j)}(t)\delta_i(N-(s_i-1)-a_j)\Delta t +$$

$$p_{(s_i+1,a_j)}(t)(d_i+\rho_i)(s_i+1)\Delta t +$$

$$p_{(s_i,a_j)}(t)(\theta_5)\Delta t \tag{17}$$

If we obtain the transition matrix as a diagonal 3D-matrix, we can show that Equation (17) as an inner product, $p(\Delta t) = P(\Delta t)\, p(0)$, where $P(\Delta t)$ is as a transition matrix and also $p(0)$ explains initial probability vector $\left( p_{(s_i+1,a_j-1)}(0), p_{(s_i,a_j+1)}(0), p_{(s_i-1,a_j)}(0), p_{(s_i+1,a_j)}, p_{(s_i,a_j)}(0) \right)$. Therefore, Equation (17) can be expressed in matrix and vector notation as:

$$p(t+\Delta t) = P(\Delta t)\, p(t) = P^{n+1}.p(0) \ , \text{ where } t = n\Delta t. \tag{18}$$

Moreover, we can derive the mean values of each susceptible $\mathbb{S}_i$ and informed $\mathbb{A}_j$, according to Equation (17). We can define them as below:

$$\begin{cases} E\big(\mathbb{S}_i(t)\big) = \displaystyle\sum_{j=0}^{m} \sum_{s=0}^{s_i} s\, p_{(s,a_j)}(t) \\[2mm] E\big(\mathbb{A}_j(t)\big) = \displaystyle\sum_{i=0}^{m} \sum_{a=0}^{a_j} a\, p_{(s_i,a)}(t) \end{cases} \tag{19}$$

### 3.5. The Variable Population Size

So far, all of the assumptions are based on a constant size of population. But, we can investigate to variation population size in calculations. To do this, it is possible to relate both rates of birth and death of population size, $N$. For this reason, we would use logistic equation, which it is proposed about population growth by Verhast [48]. This equation redefines as following:

$$\frac{dN}{dt} = b(N) - d(N) = \rho N\left(1 - \frac{N}{K}\right) \tag{20}$$

Equation (20) has shown the relation between birth, death rates and population size, N. Instead of using fix values of birth and death rated in Equation (1), we define these variables $b(N) = \rho N \text{ and } d(N) = \dfrac{\rho N^2}{K}$ and then recalculate all our calculations.

In Equation (20), $K$ and $\rho$ play important roles to increase and decrees of population size, whereas growth of K, lead to increase of death rate and the amount of people will be reduced, and vice versa. Also, we can have similar deduction for parameter $\rho$.

## 4. EXPERIMENTAL RESULTS

In order to evaluate our parameters, which were extracted from the proposed model, the assumptions of the evaluated parameters should be presented at first. All assumptions to these evaluations are the same as those mentioned earlier in section 3.1. Following steps illustrate our simulation process:

    Step 1: Deliberation of user's deterministic behaviors.

    Step 2: Compare between users behavior in the case of deterministic model and stochastic DTMC model.

    Step 3: Deliberation of threshold value affection in the extinction of information from a population.

    Step 4: Deliberation of model, according to variable size of the population.

## 4.1 The Simulation Method

Classical epidemic models have been surveyed as many literatures. The researchers of them considered those models from different views as mentioned in related work. Almost their simulator can be found in resources, which we have study most of them. However, we propose a new model of information diffusion, that is based on epidemic spreading, especially SIRS model. Moreover, we extract some parameters, that their calculation is different to related models. So, we have to implement this method by ourselves. This method is implemented with MATLAB simulator.

First step of this simulation belongs to solve of differential equations with initial value parameters, which is given in table 2. After that, there are used a discrete-time method to simulate users behavior according to the proposed model and all the transitions are based on the rates that defined between states. All these predefined values are brought in table 2. To determine transition probability, first we should calculate these probabilities at a Δt (sufficiently small) epoch, which is based on the define probabilities (rates) on move direction on the model; then we use these probabilities to conclude the probability at t+Δt time. Growth of Mean values of infected and susceptible users in comparison of deterministic mode, would be verified our approach. All other initial values and assumptions, that applied to the simulation summarized in table2, where bring below:

**Table 2.** Initial values of the parameters during the simulation

| Parameter | Description | Value |
|---|---|---|
| $N$ | Number of population | 100 |
| $b_i$ | Birth rate at each group i | 0.01 |
| $d_i$ | death rate at each group i | 0.01 |
| $m$ | Number of predefined clusters | 2 |
| $\rho_i$ | Migration rate from $S_i$ to $D_i$ | 0.2 |
| $\delta_i$ | Migration rate from $D_i$ to $S_i$ | 0.03 |
| $\varphi_i$ | Migration rate from $A_i$ to $D_i$ | 0.03 |
| $\varepsilon_i$ | Susceptibility of user in each $S_i$ groups | $\varepsilon_1 = 0.4, \varepsilon_2 = 0.6$ |
| $\gamma_j$ | Capability of transfer information for each $A_j$ groups | $\gamma_1 = 0.4, \gamma_2 = 0.7$ |
| $A_i^0$ | Initial values of each active groups $A_i$ | $A_1^0 = 20, A_2^0 = 8$ |
| $s_i^0$ | Initial values of each susceptible groups $s_i$ | $s_1^0 = 30, s_2^0 = 42$ |
| $D_i^0$ | Initial values of each deactivate groups $D_i$ | 0 |

## 4.2. Stochastic vs. Deterministic Model

This section evaluates deterministic and stochastic behaviors of the proposed model. Result of deterministic manner can be derived from the solution of differential equations, which would be considered in section 3.2. Also, as be shown in section 3.3, we could derive probabilistic transactions between states regarding to DTMC model. Then we calculate the expected number of population size at each state. To compare and verify two modes of model, we show their behavior at different time steps with $R_0 = 1.4$. This adaption will be seen in Figure 2. This figure shows the number of infected, susceptible and deactivates users of two compartments, according to time; as can be seen, all this charts have three regions, at first since the measures have been gotten in each Δt=8 msec, all six charts be linear until 10 msec (it would be like warm up phase). According to charts (a) and (b), between 10 to 20 msec we observe that activated users are increasing; because, entrance rates to these states are more than exits of it as predefined values (Table 2). After that, the trends of these charts dramatically decreased; since that, it is going to DFE state (i.e. final of epidemic); thus, the stable state would be occurred and continued. Charts (c) and (d) show behavior of susceptible users, who acts in opposite of activated users; during activated number have growth, susceptible goes down; these charts are following initial values, which have brought in Table 2; and then enter to stable state as same as previous charts. Moreover, we observe that, deactivated users grow incrementally; because, entrance rates, which initiated from active and susceptible sates, are more than exits that relates to death rate and migration to susceptible rate (see Table 2). After 40 msec, their behavior reaches to stable state, similar to other charts. Stochastic results show as same as the deterministic trend; however, in the case of stochastic mode, we observe some distortions during its progress. These deviations are happened; because, time grows with discrete form of $t = \{\Delta t, 2\Delta t, ...\}$ and at each time spaces; one event occur only, which due to calculate mean values. Before calculating these mean values, we should solve Equation (17). For this purpose, we use Equation (18), to calculate from matrix form. n in this equation depends on the expected time, t and epoch time $\Delta t$, which only one event can occur during this time. One drawback of this solution is whatever population size raise, the time and spectral complexity get higher. The major result of this complexity goes to calculation. Therefore, we should find a better solution to make less complexity. On choice is using prediction mechanism via Data mining tools. With the help of these mechanisms, such as: regression, decision tree and etc., we can forecast the end of the epidemic with amount of partial knowledge of it. One of the best benefits of these methods is their less complexity and higher speed computation in a large population size [49]. However, here we consider the epidemic model and some of the most important metrics.

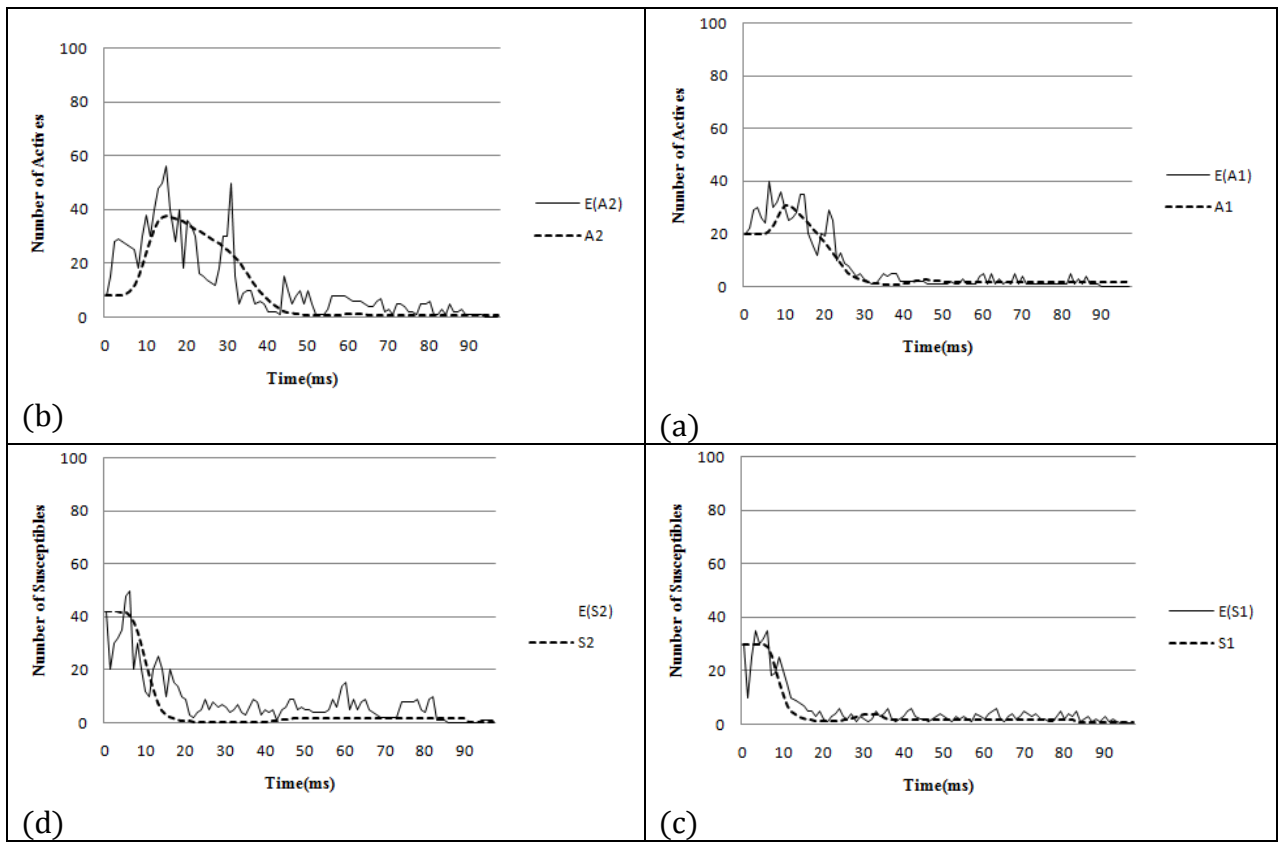

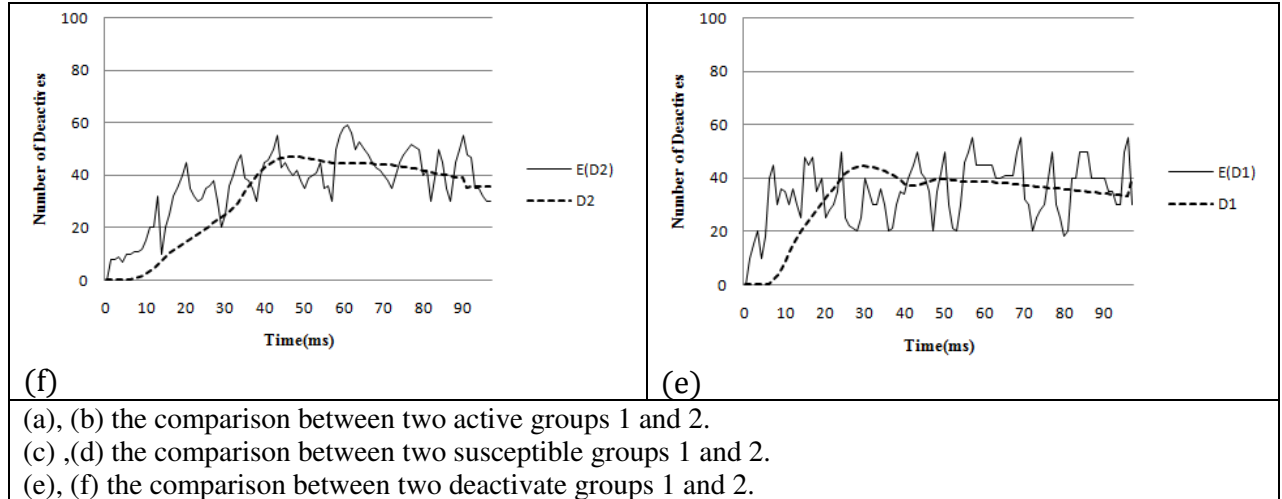

(a), (b) the comparison between two active groups 1 and 2.
(c) ,(d) the comparison between two susceptible groups 1 and 2.
(e), (f) the comparison between two deactivate groups 1 and 2.

**Figure 2.** The comparison between Deterministic (Equation (1)) and Stochastic (DTMC) behavior between susceptible, active and deactivate groups with $R_0 = 1.4$. In all charts, deterministic behavior has be shown by (*dashed curve*) and in stochastic has been shown by (*solid curve*).

## 4.3. The Impaction of on the Information Diffusion on Extinction Time

As mentioned earlier $R_0$ plays a key role on contagion. Therefore, this parameter has a significant impact on information persistence. When this value goes higher, more users get that information and take longer time to extinct. In contrast, if $R_0$ is being in low value, less users augment spreading of information and it loses its importance sooner. We get started our experiment with assign initial values to active users, according to table 2. The model is performed for variable population, 0-90. For each size of population, model has been continued until activated users go to zero. Also, by the assumption, we have two groups of activated users, so we should wait to sum of the both activated number get near zero. The time of this event stands for extinction time or the time no new information exists among the people. Impaction of parameter $R_0$ on extinction time, whole over the system is shown in Figure 3. Easily can be seen, whatever $R_0$ get increase, extinction time would be rise, too. In the case of $R_0 = 2.3$, approximately near 40msec would be our extinct time, which it has been reached to stability; because, after this time, activated number gets almost to zero. This can be observed more oblivious in Figure 2 (a) and (b). Also, for $R_0 = 4.9$, we have the same trend; however, time, which related to this value of threshold, appears in higher value, makes more interconnection between users. As a result, importance of information takes longer time to disappear.

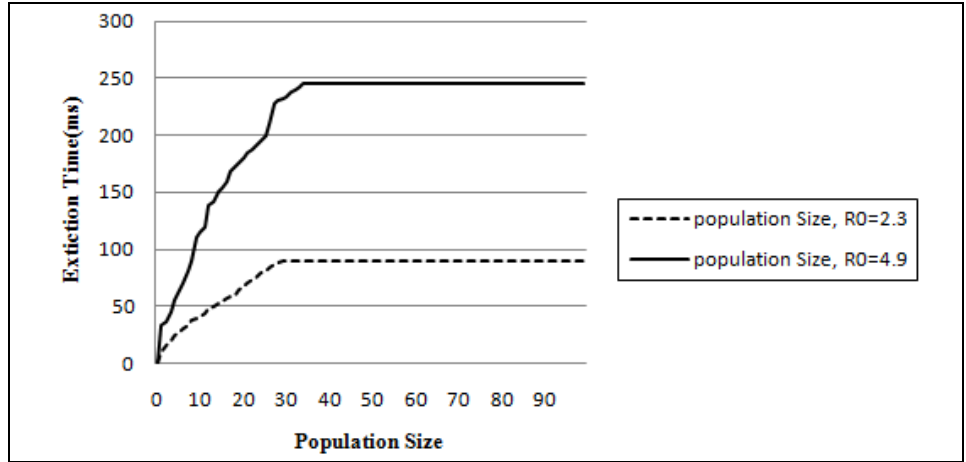

**Figure 3.** Affection of $R_0$ on the mean extinction time in whole of the system.

## 4.4. The Variable Population Size

In a social network, population dynamicity is a major feature, where can impact user behaviors'. As mentions in section 3.4, the logistic equation can model this dynamicity well. This section emphasizes on impact of parameter $K$ on the individual growth of active groups. Figure 4, shows the variation number of activated users, according to time duration. Impaction of the parameter, K calculates in two active groups, whenever this parameter is more, as a result the number of active users also raise more. The sake of this behavior refers to reduction of death rate (Equation (20)), and as follows increment of individuals in groups. This claim is justifiable in Figure 4. Also, we can see at the same value of the time, for example 20msec, number of active persons in group $A_2$, is more than $A_1$. This state come from, that susceptible user number of group $A_2$, is more than $A_1$; whose their susceptibility is in a higher place. After this time, the numbers of all activated persons get down near to zero; because, the DFE state would be started and epidemic run over among population. This state is called stable state in the system.

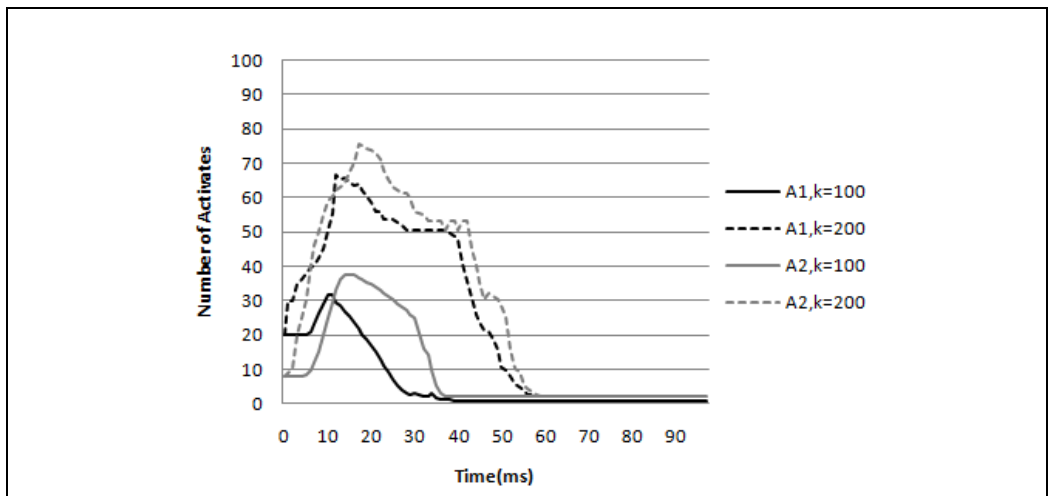

**Figure 4.** Impaction of K on the activated user rate in groups 1 and 2. $R_0 = 1.4, \rho = 1.0$.
Behavior with k = 100 is shown with (*solid curve*) and with k = 200 are shown by (*dashed curve*).

## 5. CONCLUSIONS AND FUTURE WORKS

In this paper, by the help of epidemic diseases, we proposed a general information diffusion model in social networks. Since, we grouped the users with different property, this model belongs to a compartmental model, all of users in a compartment have the same feature; but, distinct with other groups. At the first step, we have extracted a deterministic behavior of the model as a chunk differential equations, future with thank of DTMC model; stochastic behavior of the model has been deduced according to deterministic mode of the proposed model. To adaption of deterministic manner to stochastic one, we used Markovian property to calculate transition probability at a small piece of time. Then, apply this probability to obtain the mean value of population per each group. One of the important measures that we have used in this paper is a threshold measure to determine when information diffusion runs over. We extract this parameter for our model at DFE state. Moreover, the variable population size has been studied in this paper and has shown what its impaction on the number of informed users is. Also, mean of extinction time was another parameter, which extract during simulation. Relation between this parameter and threshold was depicted as a chart as well. In future work, we decide to measure the impaction of network topology on this information diffusion model and will study some important networks to show this effectiveness. As mentioned earlier, calculation related to probability of expected number, which is contained infection and susceptible users, would be more complex and has high computational order. Therefore, another study, that has worth to do, is a prediction of the number of informed users, whenever we just have initial states of users. For this reason, we can consider the growth of population as a branching process and do our calculations on this dynamic population.

## EFERENCES

[1]    A. Vázquez, R. P.Satorras and A. Vespignani, "Large-scale topological and dynamical properties of Internet," *Phys. Rev. E 65, 066130-066130-12*, Vol. 65, Issue 6, June 2002.

[2]    A. Nagurneya and Q. Qiang, "Fragile networks: identifying vulnerabilities and synergies in an uncertain age," *Wiley Intl. Trans. in Op. Res.,* Vol.19, Issue 1, March, 2010, pp.123-160.

[3]    M. Gjoka, *measurement of online social networks dissertation*, Doctor of Philosophy Thesis, Computer science Department, University of California, Irvine, 2010.

[4]    Z. Shen, K.l. Ma and T. Eliassi-Rad, "Visual Analysis of Large Heterogeneous Social Networks by Semantic and Structural Abstraction," *Visualization and Computer Graphics*, Vol. 12, Issue 6, 2006, pp.1427-1439.

[5]    E.Fee and D. M. Fox, *AIDS: The making of chronic disease,* university of California press, First edition, Berkeley, Los Angles 1992.

[6]    A.Guille and H.Hacid, "A predictive model for the temporal dynamics of information diffusion in online social networks," *WWW '12 Companion Proceedings of the 21st international,* pp.1145-1152, 2012, Lyon, France.

[7]    E.Sadikov, M.Medina, J.Leskovec and H.Garcia-Molina, "Correcting for missing data information cascades," WSDM '11 Proceedings of the fourth ACM international conference on Web search and data mining, pp. 55-64, 2011, New York, USA.

[8]    D.Kempe, J.Kleinberg and E.Tardos, "Maximizing the Spread of Influence through a Social Network," *KDD '03 Proceedings of the ninth ACM SIGKDD international conference on Knowledge discovery and data mining*, pp.137-146, 2003, Washington, USA.

[9]    H.W. Hethcote, "The mathematics of infectious diseases," *SIAM Rev*iew, Vol. 42, No. 4, 2000, pp. 599-653.

[10]   M.E.J. Newman, "The spread of epidemic disease on networks," *Phys. Rev. E 66, 016128*, Vol. 66, Issue 1, 2002.

[11]   S.Bansal, B.T. Grenfell and L.A. Meyers, "When individual behavior matters: homogeneous and network models in epidemiology," *Journal of the Royal Society Interface*, Vol. 4 No. 16, October, 2007, pp. 879-891.

[12]   H.W. Hethcote and H.R. Thieme, "Stability of the endemic equilibrium in epidemic models with subpopulations," *Mathematical Biosciences*, Vol.75, Issue 2, August, 1985, pp. 205-227.

[13]   J. M. Hyman and J. Li, *"*Differential Susceptibility and Infectivity Epidemic Models,*" Mathematical Biosciences and Engineering*, Vol. 3, No. 1, January, 2006, pp. 89-100.

[14]   X. Wang, *"*Review of Classical Epidemic Models,*"* Statistical and Applied Mathematical Sciences Institute, Undergraduate, Workshop, spring, 2010.

[15]   T. Britton, "Stochastic Epidemic Models: A survey", *Mathematical Biosciences*, Vol. 225, Issue 1, May, 2010, pp. 24-35.

[16]   J.S. Allen, *An Introduction to Stochastic Processes with Biology Applications*, First edition, Prentice Hall, 2003.

[17] J.Goldenberg, B.Libai and E.Muller, "Talk of the network: A complex systems look at the underlying process of word-of-mouth," *Marketing Letters*, springer, Vol.12, Issue 3, August 2001, pp. 211-223.

[18] K.Satio, M.Kimura, K.Ohara, and H.Motoda, "Selecting information diffusion models over social networks for behavior analysis," *Machine Learning and Knowledge Discovery in Databases, springer*, Vol. 6323, 2010, pp.180-195.

[19] M.E.J. Newman, "The structure and function of complex networks," *Society for Industrial and Applied Mathematics, SIAM Review*, Vol. 45, No. 2, 2003, pp. 167-256.

[20] D. Easley, J. Kleinberg, *Networks, Crowds, and Markets*, first edition, Cambridge University Press, Cambridge, 2010.

[21] E. Vynnycky and R.White , *An Introduction to Infectious Disease Modeling,* Oxford University Press, First edition, Oxford, 2010.

[22] J.M. Hyman and J. Li, "Differential Susceptibility Epidemic Models," *Journal of Mathematical Biology*, springer, Vol. 50, Issue 6, June, 2005, pp. 626-644.

[23] Mishra, B.K and Ansari G.M., "Differential Epidemic Model of Virus and Worms in Computer Network," *International Journal of Network Security*, Vol. 14, No. 3, May 2012, pp. 149-155.

[24] J.M Heffernan, R.J Smith and L.M Wahl, "Perspectives on the basic reproductive ratio," *J.R.Soc Interface*, Vol.2, No.4, September, 2005, pp.281-293.

[25] R. Pastor-Santorrass and A. Vespignani, "Epidemic Spreading in scale-free networks," *Phys.Rev.Lett*, Vol. 86, No. 14, 2001, pp. 3200-3203.

[26] C.Castellano and R.Pastor-Satorras, "Threshold for epidemic spreading in networks," *Phys.Rev.Let 105*, 218701, Vol.105, Issue 21, 2010.

[27] M.E.J. Newman, "Threshold Effects for Two Pathogens Spreading on a Network," *Phys. Rev. Let.95, 108701*, Vol. 95, Issue 10, 2005.

[28] D. Chakrabarti, Y.Wang, C.Wang, J. Leskovec, and C.Faloutsos, "Epidemic thresholds in real networks," *ACM Transactions on Information and System Security (TISSEC)*, Vol. 10, No.1, Issue 4, January 2008.

[29] B.A. Parkash, D. Chakrabarti, M.Faloutsos, N. Valler and C.Faloutsos, "Threshold Conditions for Arbitrary cascade models on Arbitrary Networks," *Knowledge and Information Systems, Springer*, Vol. 33, Issue. 3, December, 2012, pp. 549-575.

[30] J.A. Jacquez and P. O'Neil,"Reproduction numbers and thresholds in stochastic epidemic models. I. Homogeneous populations," *Mathematic Bioscience*, Vol.107, Issue 2, December, 1991, pp. 161-186.

[31] P. Van Den Driessche and J.Watmough, "Reproduction numbers and sub-threshold endemic equilibrium for compartmental models of disease transmission", *Mathematic Bioscience*, Vol.180, Issue 1-2, November-December 2002, pp.29-48.

[32] O.Diekmann, J.A.P. Heesterbeek and J.A.J. Metz, "On the Definition and the Computation of the Basic Reproduction Ratio R0 in Models for Infectious Diseases," *Journal of Mathematical Biology*, Vol. 28, No. 4, 1990, pp. 365–382.

[33] G.R.Fulford, M.G.Roberts and J.A.P. Heesterbeek, "The metapopulation dynamics of an infectious disease: Tuberculosis in possums," *Theoretical Population Biology*, Vol. 61, Issue 1, February, 2002, pp. 15–29.

[34] C.Fraser, S.Riley, R.Anderson and N.M. Ferguson, "Factors that make an infectious disease outbreak controllable," *proceeding the national academy of sciences of the united states of america*, Vol. 101, 2004, pp. 6146–6151.

[35] M.Boguna and R.Pastor-Satorras, "Epidemic spreading in correlated complex networks," *Phys Rev E 66, 047104*, Vol.66, Issue 4, 2002.

[36] A.N.Hill and I.M. Longini, "The critical vaccination fraction for heterogeneous epidemic models," *Mathematic Bioscience*, Vol.181, Issue 1, January, 2003, pp.85-106.

[37] Y. Wang, D.Chakrabarti, C.Wang and C.Faloutsos, "Epidemic spreading in real networks: An eigenvalue viewpoint," *Reliable Distributed Systems Proceedings 22nd International Symposium*, pp. 25–34, October, 2003, Florence, Italy.

[38] M.E.Alexander and S.M.Moghadas, "Bifurcation analysis of an SIRS epidemic model with generalized incidence," *SIAM Journal on Applied Mathematics*, Vol. 65, Issue 5, 2005, pp. 1794–1816.

[39] I.Z.Kiss, D.M.Green and R.R. Kao, "The effect of contact heterogeneity and multiple routes of transmission on final epidemic size," *Mathematic Bioscience*, Vol. 203, 2006, pp. 124–136.

[40] R. Anderson and R.M. May, *Infectious Diseases of Humans: Dynamics and Control*, Oxford University Press, Oxford, 1991.

[41] N. Masuda and N. Konno, "Multi-state epidemic processes on complex networks," *Journal of Theoretical Biology*, Vol. 243, Issue 1, November 2006, pp. 64–75.

[42] K. B. Blyuss and Y.N. Kyrychko, "On a basic model of a two-disease epidemic," *Applied Mathematics and Computation*, Vol. 160, Issue 1, January, 2005, pp. 177–187.

[43] J.A. Hyman and J. Li, "The reproductive number for an HIV model with differential infectivity and staged progression," *Linear Algebra and its Application*, Vol. 398, March, 2005, pp.101–116.

[44] M. Salmani and P.van den Driessche, "A model for disease transmission in a patchy environment," *Discrete Continuous Dynamical Systems Series B*, Vol. 6, No. 1, 2006, pp. 185–202.

[45] L. Zager, "Thresholds infection processes on networks with structural uncertainly," Ph.D Thesis, EECS Dept., Massachusetts institute of technology, 2008.

[46] D.S. Bernstein, *Matrix Mathematics: Theory, Facts, and Formulas*, second edition, Princeton University, New Jersey, 2009.

[47] W.K. Ching and M.K. Ng, *Markov Chains: Models, Algorithms and Applications*, First edition, France, Springer, 2006.

[48] N. Bacaer, *A Short History of Mathematical Population Dynamics*, First edition, London, Springer, 2011.

[49] A.Nagar, L.Denoyer and P.Gallinari, "Predicting Information Diffusion on Social Networks with Partial Knowledge," *WWW 2012 Companion*, pp. 16-20, April, 2012, Lyon, France.